# Designing Magnetic Droplet Soliton Nucleation Employing Spin Polarizer

Morteza Mohseni[1, §] and Majid Mohseni[1,*]

[1]*Faculty of Physics, Shahid Beheshti University, Evin, 19839 Tehran, Iran*

**Abstract**

We show by means of micromagnetic simulations that spin polarizer in nano-contact spin torque oscillators (NC-STOs) as the representative of the fixed layer in an orthogonal pseudo-spin valve (P-SV) can be employed to design and to control magnetic droplet soliton nucleation and dynamics. We found that using a tilted spin polarizer layer decreases the droplet nucleation time which is more suitable for high speed applications. However, a tilted spin polarizer increases the nucleation current and decreases the frequency stability of the droplet. Additionally, by driving the magnetization inhomogenously at the nano-contact region, it is found that a tilted spin polarizer reduces the precession angle of the droplet and through an interplay with the Oersted field of the DC current, it breaks the spatial symmetry of the droplet profile. Our findings explore fundamental insight into nano-scale magnetic droplet soliton dynamics with potential tunability parameters for future microwave electronics.

[§] Current address: Fachbereich Physik and Landesforschungszentrum OPTIMAS, Technische Universität Kaiserslautern, 67663 Kaiserslautern, Germany
[*] Corresponding author's email address: m-mohseni@sbu.ac.ir, majidmohseni@gmail.com

Nano-scale self-localized standing spin wave solitons, from a large family of magnetic solitons are a subject of interest both from physical and technological points of view [1-5]. Formation and control of these types of solitons at the nano-scale circumvents challenges with advanced materials and device architectures. For example, in current driven spin transfer torque (STT) devices, contribution of different mechanisms, e.g. the STT effect, Gilbert damping parameter, exchange interaction and magnetic anisotropy are known to influence on nucleation and dynamics of these solitons significantly [6-10]. On the other hand, such magnetic objects have introduced a new paradigm in the field of nano-magnetism where they can be applied in information processing technologies, synapse architectures and magnetic storages [11-15]. In addition and more interestingly, due to their dynamical response and frequency of operation, they can be considered for relevant microwave applications, e.g. nano-contact spin torque oscillators (NC-STOs) and maybe later in neuromorphic computing systems. Furthermore, it can be noted that the oscillation amplitude of a magnetic soliton in NC-STOs is higher than other types of oscillators mainly due to the localization effect with large spin precession angle [16,17].

Up to now, different magnetic solitons have been studied and observed. Namely skyrmions [12], vortices [18], bullets [1,19] and droplets [20,21]. Droplets were suggested in 2010 to be observed in a current driven NC-STOs with a large perpendicular magnetic anisotropy (PMA) in the free layer [10] as a dissipative counterpart of magnon drops which was introduced for the first time in 1977 [22]. Droplet as a localized dynamical excitation can be formed beneath the nano-contact (NC) when the STT of the applied current balances the damping and the PMA (non-linearity) is acting against the dispersion. It was predicted that the magnetization of the droplet is almost completely reversed relative to the magnetization of the film outside the NC perimeter [10]. This kind of soliton with large amplitude of oscillation has been envisioned as a distinguishable candidate for high amplitude microwave oscillators and rapid information processing devices.

The initial theory of the droplet considered the magnetization of the fixed layer in a spin valve (SV) to be completely out-of-the film plane [10]. Nevertheless, first experimental studies showed that droplets can be nucleated in NC-STOs with orthogonal pseudo-SVs (P-SVs) structures utilizing a fixed layer with tilted magnetization in the presence of an out-of-plane (OOP) field [20]. Indeed, the employed orthogonal P-SV consisted of a Co layer as the spin polarizer (fixed layer) with in-plane magnetic anisotropy and a Co/Ni multilayer with a strong PMA as the free layer. Following studies indicated that droplets can be nucleated in NC-STOs employing other type of fixed layers, e.g. permalloy (Py), with a smaller saturation magnetization than the Co layer [21]. Subsequent to first observations, additional works illustrated that the magnetization of the fixed layer plays a major role on the droplet nucleation boundaries, mainly due to an interplay among different components of the STT and the STT asymmetry [23]. It is worth to mention that, since it is crucial for the free layer to benefit from a strong PMA, a Co/Ni multilayer is one of the most promising candidates [24,25] that used in all those mentioned devices.

By far, all successful experiments about the droplet were performed in NC-STOs in which they seem to be different only in their fixed layers. However, their results possess inconsistencies from certain points of view. One of the conflicts refers to the frequency shift of the droplet with respect to the applied current. Although droplet frequency in NC-STOs with Co fixed layer exhibited a red shift (decrease of the frequency by increasing the current amplitude) [20], devices that employed Py as the fixed layer displayed a blue shift [21], which was not in accordance with predictions [10].

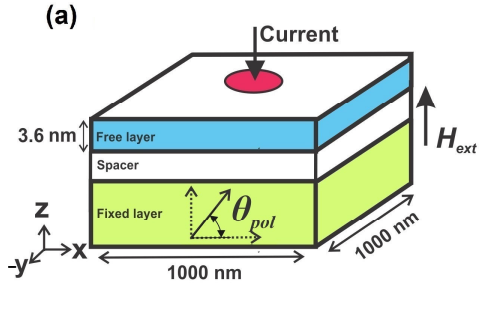 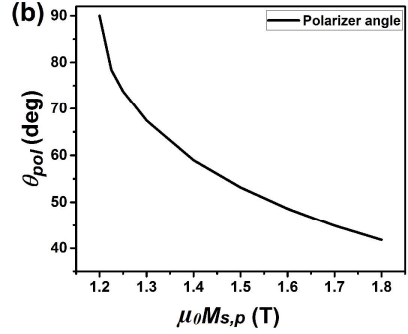

Figure 1. a) A Schematic picture of the device structure, b) Calculated spin polarizer angle $\theta_{pol}$ with respect to saturation magnetization of the fixed layer $M_{s,p}$, while the external field is fixed at $\mu_0 H_{ext} = 1.2\, T$.

Another inconsistency can be referred to the precession angle of the droplets, where it has been observed to be different in orthogonal [26] and all-perpendicular NC-STOs [27]. Achieving stable droplet in NC-STOs with a Py fixed layer in comparison to the less stable one with a Co fixed layer can be counted as another inconsistency that needs to be understood.

However, in addition to those aforementioned points, the role of the fixed layer on the spatial profile, nucleation time and frequency stability of the droplet have never been investigated in details. The aim of this study is to address some of these questions and to provide a deeper insight into the droplet nucleation and dynamics considering the spin polarizer layer. On the other hand, our findings should stimulate further interests in studying configurability and stability of the droplet via possible and accessible mechanisms to make them more suitable for real applications.

We performed micromagnetic simulations utilizing GPU-based MuMax 3.0 [28] open source software, considering the free layer of the device as the simulated geometry. The magnetization dynamics of the system calculated based on the Landau-Lifshitz-Gilbert-Slonczewski equation [28,29]

$$\frac{\partial \hat{m}}{\partial t} = -\gamma \hat{m} \times \mu_0 \vec{H}_{eff} + \alpha \hat{m} \times \frac{\partial \hat{m}}{\partial t}$$
$$-\gamma \mu_0 M_s \sigma(I) \varepsilon \hat{m} \times (\hat{m} \times \hat{M}) \quad (1)$$

where $\vec{H}_{eff}$ and $\alpha$ are the effective field and the Gilbert damping factor of the free layer, respectively. $\gamma/2\pi = 28\, GHz/T$ is the gyromagnetic ratio, $\hat{m}$ and $\hat{M}$ are the magnetizations of the free and fixed layer, respectively. $M_s$ is the saturation magnetization of the free layer. $I$ indicates the applied current, $\sigma(I)$ is the dimensionless spin torque coefficient and $\varepsilon$ refers to the spin torque efficiency.

It is possible to achieve a canted spin polarizer layer via applying an in-plane field in an all-perpendicular NC-STOs or an OOP field in an orthogonal NC-STO which we would like to undertake the later configuration in our study. A schematic image of the device under this study is presented in Fig. 1a. The free layer is assumed to have a dimensions equal to 1000×1000×3.6 nm (3.6 nm is thickness of the free layer), which is divided into 256×256×1 cells that gives cell sizes which are below the exchange length. The material parameters which are used for the simulations are taken from Ref. [20] and are as follows: $\alpha = 0.03$, $A_{exch} = 12.5\, pJ/m$, $M_s = 737\, kA/m$, uniaxial PMA $K_u = 443\, kJ/m^3$, polarization $P=0.43$ and spin torque asymmetry $\lambda = 1.3$. Absorbing boundaries are considered to avoid back-reflections and the role of the temperature is not taken into account. The DC current which is incorporated with the effect of the Oersted field (unless otherwise is mentioned), assumed to pass through a cylinder-like NC with different diameters in the middle of the system with negative polarity.

In order to find the spin polarization angle $\theta_{pol}$ (which is defined via the magnetization angle of the fixed layer), the magnetostatic boundary conditions of the fixed layer can be considered. This can be derived from Maxwell equations for magnetic fields [30] in the studied system, as

$$\vec{B}_e \cdot \hat{n} = \vec{B} \cdot \hat{n}$$
$$\vec{H}_e \times \hat{n} = \vec{H} \times \hat{n} \quad (2)$$

where it indicates that the normal component of the magnetic flux, and the tangential component of the magnetic field are continuing parameters at the interface between two media (inside and outside the layer).

At the first part of this study, we assume that the saturation magnetization of the fixed layer $M_{s,p}$, varies from $\mu_0 M_{s,p} = 1.2\ T$ to $\mu_0 M_{s,p} = 1.8\ T$. By applying a perpendicular external field with an amplitude of $\mu_0 H_{ext} = 1.2\ T$ and via solving the Equ. 2 numerically, we can find the magnetization angle (defined with respect to the film plane) of the fixed layer $\theta_{pol}$, which is plotted in Fig. 1b. As it is shown in Fig. 1b, increasing the saturation magnetization of the fixed layer, decreases its angle from an all perpendicular $\theta_{pol} = 90°$ to a tilted $\theta_{pol} = 41°$ orientation. We consider the calculated spin polarizer angle presented in Fig. 1b in the simulations (unless otherwise is mentioned) to find the influence of this parameter on the droplet nucleation and dynamics. At this part of the study, we set the diameter of the NC to $100\ nm$.

The influence of the spin polarizer angle on the nucleation frequency of the droplet is plotted in Fig. 2a. We define the maximum frequency of the droplet, $f_{max}$, to be the frequency at the nucleation current, considering that the droplet has a red shift trend with the applied current. It is clear that $f_{max}$ of the droplet is constant at $f_{max} = 34.5\ GHz$ regardless of any changes in the $\theta_{pol}$. This can be explained by considering the fact that, damping-like STT is required for the droplet to sustain beneath the NC, which does not affect its frequency [10,20,21]. On the other hand, increasing (decreasing) the $M_{s,p}$ ($\theta_{pol}$), increases the droplet nucleation current $I_N$, as shown in Fig. 2b. Our numerically obtained results which are in accordance with the reported measurements and calculations [20, 21, 23], declare that a lower nucleation current can be achieved via fixed layers with smaller saturation magnetization, e.g. for Py compared to Co. We argue the existence of this trend as an indication of the reduced spin polarization by increasing the $M_{s,p}$, since in higher (lower) $M_{s,p}$ ($\theta_{pol}$), the polarization of the current is mainly driven by the z-component of the spin polarizer which is mainly required for droplet nucleation [10,23]. These results suggest that optimizing the required energy for the droplet nucleation is possible to be configured via material parameters of the fixed layer, even if the applied field remains fixed. Indeed, it is favourable to choose materials with smaller saturation magnetization as the fixed layer since the frequency remains the same while a lower operational current can be achieved.

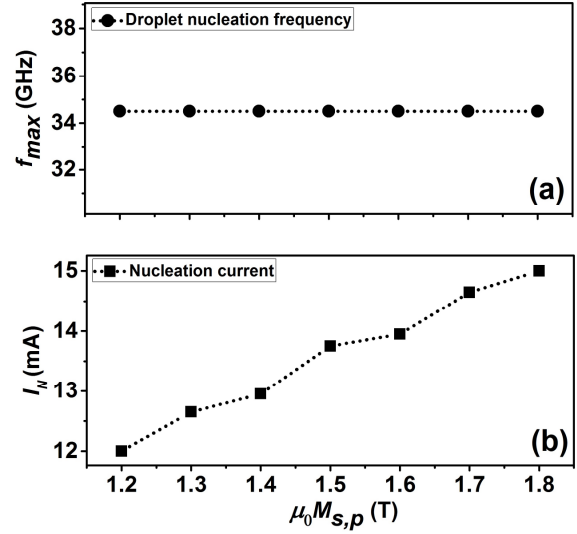

Figure 2. a) Droplet nucleation frequency and, b) droplet nucleation current with respect to $\mu_0 M_{s,p}$.

One of the remaining questions that need to be answered is the nucleation time of the droplets. Fig. 3a shows the time evolution of the OOP component of the free layer underneath the NC in NC-STOs with different spin polarizers ($\mu_0 M_{s,p} = 1.2\ T$ and $\mu_0 M_{s,p} = 1.8\ T$). The applied current is set to $I = 20\ mA$ for both devices. It is clear that the nucleation time of the droplet for a tilted polarizer ($\mu_0 M_{s,p} = 1.8\ T$) is shorter in comparison to the perpendicular polarizer ($\mu_0 M_{s,p} = 1.2\ T$). We define the nucleation time of the droplets, $t_N$, as the moment when instability underneath the NC starts to appear and the magnetization begins to reverse, as depicted in Fig. 3a. Furthermore, the nucleation time of the droplet with respect to the applied current for both devices is displayed in Fig. 3b. Increasing the current, decreases the nucleation time since a higher STT applies to compensate the damping, based on Equ. 1. However, using a canted polarizer ($\mu_0 M_{s,p} = 1.8\ T$) decreases the $t_N$ significantly. This can be explained based on a schematic diagram of the fixed and the free layer magnetizations before $t_N$, as shown in Fig. 3c.

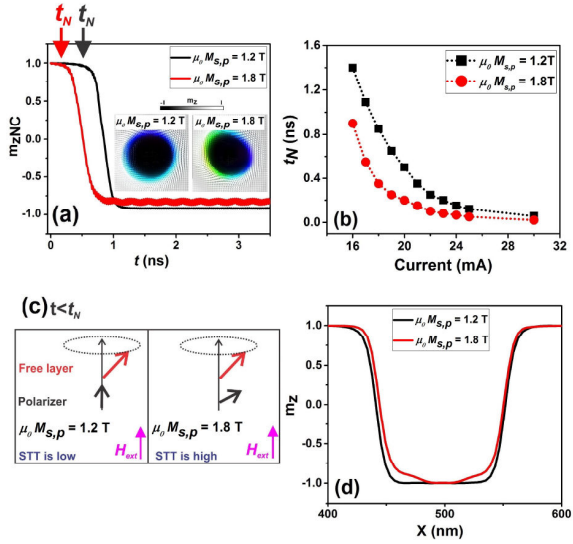

Figure 3. a) Time evolution of the OOP component of the magnetization underneath the NC $m_{zNC}$ and the defined nucleation time $t_N$. Note that the current is applied at $t = 0$, b) Droplet nucleation time with respect to the applied current, c) Magnetization configuration of the fixed and the free layers and their relevant STT for perpendicular ($\mu_0 M_{s,p} = 1.2\ T$) and tilted polarizers ($\mu_0 M_{s,p} = 1.8\ T$) before the droplet nucleation, d) Spatial distribution of the OOP component of the free layer magnetization $m_z$, extracted along X (at Y=0, beneath the NC). Inset of (a) shows a snapshot of the droplet at $t = 3.5$ ns. Note the differences in $M_{s,p}$ in all figures. $I = 20\ mA$ for (a) and (d).

Indeed, before $t_N$ when the free layer is at the normal precession state, the STO device with the $\mu_0 M_{s,p} = 1.8\ T$ represents a rather larger angle between the fixed and the free layer magnetizations than the STO device with the $\mu_0 M_{s,p} = 1.2\ T$. Hence, in the former one, more STT applies to the free layer than the later one in order to make the free layer modulationally unstable (considering Equ .1). Therefore, droplet forms within a shorter time in STO device with $\mu_0 M_{s,p} = 1.8\ T$.

One may find an inconsistency between the nucleation time and the nucleation current. Here, as the $M_{s,p}$ increases, the nucleation current increases and nucleation time decreases. We emphasize that there are two different regimes we discuss here. The nucleation current of the droplet is mainly governed by the z-component of the $M_{s,p}$. While the nucleation time is governed by the relative angle between the fixed and the free layers and addresses how fast modulational instability occurs before droplet formation. Based on Ref. [23], by increasing the $M_{s,p}$, an increase in $I_N$ is expected. However, relying on Equ .1, before the droplet nucleation, highest STT can be achieved when the magnetization of the free and fixed layers are orthogonal to each other which decreases the nucleation time [31].

The time evolution of the OOP component of the magnetization beneath the NC ($m_{zNC}$) represented in Fig. 3a, addresses another interesting fact. By taking into account the absolute value of the $m_{zNC}$ of the droplet, one can find that although for the perpendicular spin polarizer the magnetization beneath the NC is completely reversed, presence of a tilted spin polarizer results in a partially reversed magnetization. This can be realized more explicitly in Fig. 3d, which presents the spatial distribution of the OOP component of the magnetization of the free layer $m_z$ along X (at Y = 0). This figure confirms a slight reduction of the magnetization reversal underneath the NC when polarizer is tilted strongly ($\mu_0 M_{s,p} = 1.8\ T$).

This can provide a possible explanation to recent observations where the magnetization of the droplet with a tilted fixed layer was observed to be partially reversed [26], but for an all-perpendicular NC-STOs was observed to be almost completely reversed [27]. In addition, this might explain why droplet has shown to be more stable for the devices with Py as the fixed layer [21] than that with the Co one [20], in the presence of an OOP field of approximately $\mu_0 H_{ext} = 1\,T$ when Py is completely saturated but Co is tilted.

The inset of the Fig. 3a shows a snapshot of the droplet at $t = 3.5\,ns$ for two STO devices with different fixed layers. The spatial profile of the droplet with a perpendicular polarizer ($\mu_0 M_{s,p} = 1.2\,T$) is fully symmetric whereas this is not similar to the tilted polarizer ($\mu_0 M_{s,p} = 1.8\,T$). In order to have a better illustration of the spatial profile of the droplet, a Fast Furrier Transform (FFT) of the droplet dynamics with different spin polarizer angles of $\theta_{pol} = 90°$ ($\mu_0 M_{s,p} = 1.2\,T$), $\theta_{pol} = 53.1°$ ($\mu_0 M_{s,p} = 1.5\,T$) and $\theta_{pol} = 41°$ ($\mu_0 M_{s,p} = 1.8\,T$) when NC is driven by $I = 20\,mA$ are plotted in Fig. 4a-c. As the angle of the spin polarizer decreases from Fig. 4a to Fig. 4c, the spatial profile of the droplet becomes more asymmetric. It seems that the in-plane component of the spin polarizer tends to drive the magnetization beneath the NC inhomogeneously which finally leads to an asymmetric droplet profile. The inhomogeneity of the droplet depends on the direction of the spin polarizer, as can be illustrated by comparing Fig. 4c with Fig. 4d. More importantly, the role of the Oersted field of the applied current should be taken into account since it can break the spatial symmetry of the effective field around and inside the NC [19, 32]. Considering the droplet, presence of a tilted spin polarizer together with the Oersted field of the applied current can break the spatial symmetry of the system and hence, the droplet dynamics is not spatially homogenous anymore.

The role of the Oersted field as a source of field inhomogeneity around and inside the NC

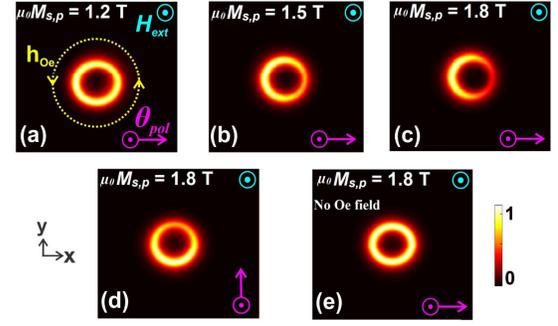

Figure 4. a) FFT (10 ns) of the droplet oscillations for a) $\mu_0 M_{s,p} = 1.2\,T$, b) $\mu_0 M_{s,p} = 1.5\,T$, c) $\mu_0 M_{s,p} = 1.8\,T$, d) $\mu_0 M_{s,p} = 1.8\,T$ when the direction of the spin polarizer is set differently and, e) $\mu_0 M_{s,p} = 1.8\,T$ when the Oersted field of the DC current is set to zero. $I = 20\,mA$ and $\mu_0 H_{ext} = 1.2\,T$ for all figures. Note the direction of the applied field (green colour coded) and the direction of the spin polarizer (purple colour coded).

is major in such STO devices, since the spatial profile of the droplet is fully symmetric without the presence of the Oersted field, even if the spin polarizer is canted. This can be confirmed from Fig. 4e.

Up to this point, we discussed the droplet dynamics when the NC diameter is fixed. However, it is known from the theory of the droplet that when the polarizer is fixed, the nucleation current depends on the spin torque asymmetry and the NC radius [10].

Fig. 5a shows the droplet nucleation current density with respect to the NC diameter for the certain polarizations of $\mu_0 M_{s,p} = 1.2\,T$ and $\mu_0 M_{s,p} = 1.8\,T$ (external field is fixed at $\mu_0 H_{ext} = 1.2\,T$). Our evidences which are in agreement with the theory of the droplet [10], shows that increasing the NC diameter decreases the required current for droplet nucleation. However, here we see that employing a perpendicular polarizer ($\mu_0 M_{s,p} = 1.2\,T$) reduces the nucleation current significantly. For example, for a NC with a diameter of $20\,nm$, the difference between the nucleation current density for the perpendicular ($\mu_0 M_{s,p} = 1.2\,T$) and tilted ($\mu_0 M_{s,p} = 1.8\,T$) polarizers is equal to:

$$\Delta J_N = J_N(M_{s,p,1.2\,T}) - J_N(M_{s,p,1.8\,T}) = 6.4 \times 10^{12}\,Am^{-2}.$$

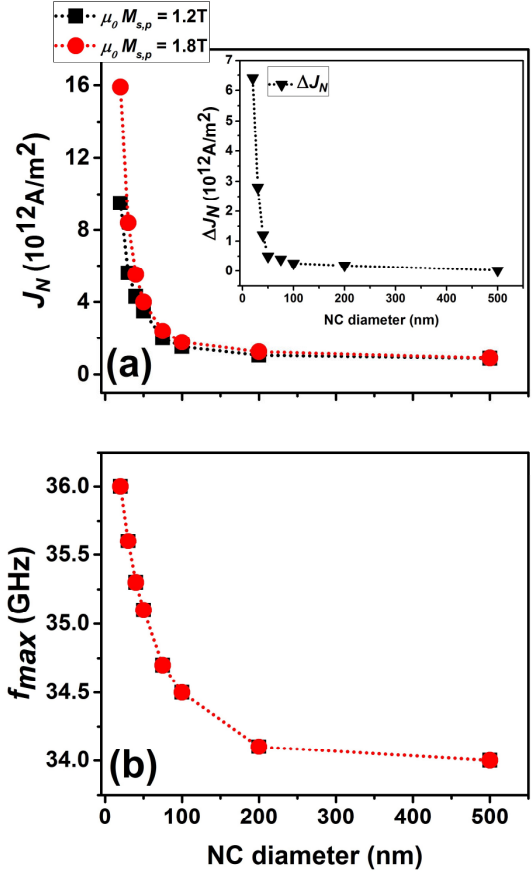

Figure 5. a) Droplet nucleation current density and, b) droplet maximum frequency with respect to the NC diameters. The inset of (a) shows $\Delta J_N = J_N(M_{s,p,1.2\,T}) - J_N(M_{s,p,1.8\,T})$ with respect to the NC diameters.

$\Delta J_N$ also decreases by increasing the NC diameter, as depicted as the inset of Fig. 5a. The existing difference in the nucleation current of the two different spin polarizers (with respect to the NC diameters) can be explained by the fact that increasing the $M_{s,p}$, decreases the spin polarization of the applied spin torque (refer to the z-component), and for the droplet nucleation, this should be compensated by an increase in the current density. In fact, nucleation current balances the damping which allows the droplet to sustain beneath the NC, considering the interaction between PMA (non-linearity) and exchange (dispersion).

NC diameter is also effective on the droplet frequency [10, 23]. The nucleation frequency with respect to the NC diameters is plotted in Fig. 5b. The $f_{max}$ decreases approximately by 2 $GHz$ when the NC diameter increases from 20 $nm$ to 500 $nm$. This trend is qualitatively in agreement with earlier predictions [10]. As mentioned earlier in Fig. 2a, spin polarizer does not change the droplet frequency, which again confirmed in Fig. 5b, since the frequency of the droplet for $\mu_0 M_{s,p} = 1.2\,T$ and $\mu_0 M_{s,p} = 1.8\,T$ are identical if the NC diameter is fixed.

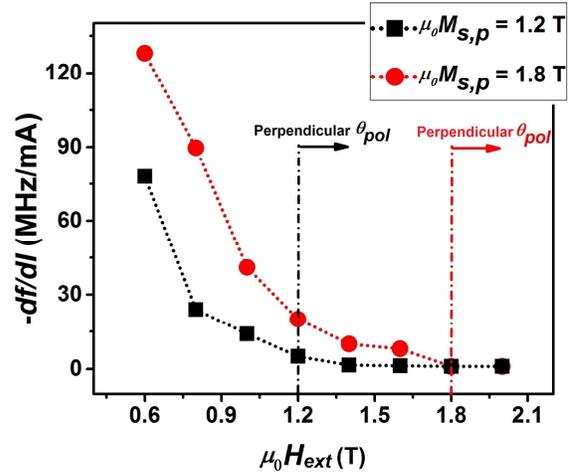

Figure 6. Current frequency tunability (-df/dI) of the droplet with respect to the applied field for $\mu_0 M_{s,p} = 1.2\,T$ and $\mu_0 M_{s,p} = 1.8\,T$.

We briefly addressed at the introduction that the current frequency tunability of the droplet were inconsistent in experiments, as reported in previous observations [20, 21]. Our results are still in accordance with the theory of droplet and some of experimental observations and exhibit a frequency red shift response. Although in our NC-STOs, we can see that how the spin polarizer affects the frequency response of the droplet. Current swept frequency tunability of the droplet, -df/dI, for STOs with $\mu_0 M_{s,p} = 1.2\,T$ and $\mu_0 M_{s,p} = 1.8\,T$ against the applied field is presented in Fig. 6. It is obvious that increasing the applied field (which means an increase in the spin polarizer angle $\theta_{pol}$, to become more perpendicular to the film plane), decreases the absolute value of the -*df/dI* and makes the frequency of the droplet to become unchanged versus field and current. The -*df/dI* decreases down to approximately 1 $MHz/mA$ if the fixed layer become completely perpendicular. However, employing a fixed layer with smaller saturation magnetization ($\mu_0 M_{s,p} = 1.2\ T$), decreases the -*df/dI* more pronouncedly than that with $\mu_0 M_{s,p} = 1.8\ T$. This result again confirms that using smaller saturation magnetization fixed layers allows the frequency of operation to be less influenced by external parameters, e.g. current and field.

In summary, our results reveal that spin polarizer which indeed is the representative of the fixed layer in an orthogonal NC-STOs, can be employed to design and to configure droplet nucleation and dynamics. We found that, using a tilted spin polarizer increases the nucleation current, reduces the angle of precession and decreases the frequency stability of the droplet. However, using a tilted fixed layer benefits from a faster nucleation time. Our finding are interesting not only to understand nano-scale magnetic solitons and spin wave dynamics in NC-STOs, but also can be employed in designing and optimizing fast electronics and microwave systems utilizing magnetic solitons. In addition, they should stimulate further interests in studying configurability of spin wave solitons in NC-STOs to make them more feasible for real applications.

M. Mohseni acknowledges support from Iran Science Elites Federation (ISEF), Iran National Science Foundation (INSF), Iran Nanotechnology Initiative Council (INIC) and Iran's National Elites Foundation (INEF). Fruitful discussions with H. F. Yazdi and M. Hamdi from Shahid Beheshti University is appreciated.